\documentclass[]{spie}  

\usepackage{amsmath,amsfonts,amssymb}
\usepackage{graphicx}
\usepackage[colorlinks=true, allcolors=blue]{hyperref}

\title{Optimizing the Efficiency of Fabry-Perot Interferometers with Silicon-Substrate Mirrors}

\author[a]{Nicholas F. Cothard}
\author[b]{Mahiro Abe}
\author[c]{Thomas Nikola}
\author[d]{Gordon J. Stacey}
\author[c]{German Cortes-Medellin}
\author[b]{Patricio A. Gallardo}
\author[b]{Brian J. Koopman}
\author[b]{Michael D. Niemack}
\author[d]{Stephen C. Parshley}
\author[b]{Eve M. Vavagiakis}
\author[b]{Kenneth J. Vetter}
\affil[a]{Department of Applied and Engineering Physics, Cornell University, Ithaca, NY 14853, USA}
\affil[b]{Department of Physics, Cornell University, Ithaca, NY 14853, USA}
\affil[c]{Cornell Center for Astrophysics and Planetary Sciences, Ithaca, NY 14853, USA}
\affil[d]{Department of Astronomy, Cornell University, Ithaca, NY 14853, USA}

\authorinfo{Further author information: (Send correspondence to N.F.C.)\\N.F.C.: E-mail: nc467@cornell.edu, Telephone: 1 607 255 0833}

\begin{document} 
\maketitle

\begin{abstract}
We present the novel design of microfabricated, silicon-substrate based mirrors for use in cryogenic Fabry-Perot Interferometers (FPIs) for the mid-IR to sub-mm/mm wavelength regime. One side of the silicon substrate will have a double-layer metamaterial anti-reflection coating (ARC) anisotropically etched into it and the other side will be metalized with a reflective mesh pattern. The double-layer ARC ensures a reflectance of less than 1\% at the surface substrate over the FPI bandwidth. This low reflectance is required to achieve broadband capability and to mitigate contaminating resonances from the silicon surface. Two silicon substrates with their metalized surfaces facing each other and held parallel with an adjustable separation will compose the FPI. To create an FPI with nearly uniform finesse over the FPI bandwidth, we use a combination of inductive and capacitive gold meshes evaporated onto the silicon substrate. We also consider the use of niobium as a superconducting reflective mesh for long wavelengths to eliminate ohmic losses at each reflection in the resonating cavity of the FPI and thereby increase overall transmission. We develop these silicon-substrate based FPIs for use in ground (e.g. CCAT-prime), air (e.g. HIRMES), and future space-based telescopes (e.g. the Origins Space Telescope concept). Such FPIs are well suited for spectroscopic imaging with the upcoming large IR/sub-mm/mm TES bolometer detector arrays. Here we present the fabrication and performance of multi-layer, plasma-etched, silicon metamaterial ARC, as well as models of the mirrors and FPIs. 

\end{abstract}

\keywords{Fabry-Perot interferometer, etalon, metal mesh, silicon substrate, metamaterial, anti-reflection coating, deep reactive ion etching}

\section{INTRODUCTION}
\label{sec:intro}  
Fabry-Perot Interferometers (FPIs) are used in many spectrometers where throughput (Jacquinot advantage) and spatial multiplexing is important. In the infrared to millimeter regime, cryogenic scanning FPIs have been used within many cutting-edge astrophysical instruments including ground-based mid-IR \cite{lacy-thesis} and sub-mm \cite{bradford} imaging FPIs, airborne far-IR single beam and imaging FPIs \cite{watson-thesis,poglitsch,latvakoski} and in the Infrared Space Observatory \cite{clegg}. A FPI consists of two highly reflective surfaces held parallel to each other, forming a high quality factor resonance cavity known as an etalon. To date, all FPIs used for astrophysics in the far-IR to sub-mm bands have relied on mirrors made from free-standing metal mesh which are stretched on stainless steel or nickel metal rings. Metal meshes behave as frequency dependent filters for wavelengths longer than the mesh's grid constant (pitch). A mesh of horizontal and vertical metal wires acts as a high-pass filter and is known as an inductive mesh. An array of metal patches acts as a low-pass filter and is known as a capacitive mesh but these cannot be fabricated without a mesh substrate. The filter response of a mesh is well controlled by adjusting the geometry of the mesh elements. Etalons made with gold flashed free-standing inductive meshes can achieve typical finesses of $\sim\!50$ and an overall transmission of $\sim70\%$.

The mirrors discussed here will be lithographically patterned metal meshes on a silicon substrate. Silicon-substrate based mirrors promise significant improvements over free-standing metal meshes in bandwidth and mechanical stability. Free-standing meshes have a very small periodicity ($\sim \lambda/3$) and are therefore very fragile. Strength can be increased by enlarging their wire thickness but this increases their reflectivity and absorptivity, degrading their overall performance. In order to meet mirror flatness requirements, free-standing meshes are typically made of gold flashed nickel in order to withstand high stresses due to stretching. The gold flashing typically has a surface roughness $\sim1\,\mu\textrm{m}$ which can cause different positions in the cavity to resonate at different wavelengths. Free-standing meshes are limited to the inductive mesh geometry which has a strongly wavelength-dependent reflectivity and therefore have a strongly wavelength-dependent finesse since $F = \pi\sqrt{R}/(1-R)$. This greatly limits the bandwidth of a free-standing metal mesh FPI since finesses less than 20 have low spectral purity and can cause inter-order leakage, while finesses greater than 70 have too little transmission due to the absorptivity of the meshes. Typical free-standing metal mesh FPIs are limited to bandwidths of $\sim\!1\!:\!1.6$ while our applications require bandwidths of $\sim\!1\!:\!2$.

The structural stability and flexible fabrication techniques of silicon-substrate based metal mesh mirrors can mitigate these issues. Metal meshes will be lithographically patterned on flat silicon wafers, eliminating the fragility and flatness issues of free-standing meshes. The crystalline structure of silicon will also limit substrate vibrations to high frequencies, even for large diameter samples. The maturity of lithographic technologies for the silicon industry will enable flexibility in the mesh design by allowing any combination of inductive and capacitive geometries. Additionally, high resistivity silicon has a low loss tangent in the far-IR and millimeter, making it an ideal substrate for high-throughput devices.

However, because of silicon's high index of refraction, the mirror substrates require an anti-reflection coating (ARC) to mitigate Fresnel reflections. Dielectric films with specific indices and thicknesses can be used as ARCs. However, their indices are difficult to control and the mismatched coefficients of thermal expansion (CTEs) between the films and silicon make them a challenging technology for cryogenic mirrors which have a strict flatness requirement. As an alternative, silicon metamaterial ARCs have been developed and successfully deployed on telescope instruments, such as ACTPol and AdvACTPol \cite{datta_large-aperture_2013}. Metamaterial ARCs consist of three dimensional sub-wavelength structures on the surface of the silicon optic. Precision micromachining techniques enable the fabrication of quarter-wavelength ARCs with an effective index determined by the micromachined geometry. Stacking multiple quarter wavelength layers provides wider bandwidth ARCs \cite{gallardo_deep_2016}.


This paper discusses the development of silicon-substrate based metal mesh mirrors for use in scanning FPIs  in upcoming ground (e.g. CCAT-prime \cite{ccat-prime,eve-ccatp-spie2018}), air (e.g. HIRMES \cite{hirmes,greg-hirmes-spie2018}), and future space-based spectrometers (e.g. NASA's Origins Space Telescope concept \cite{origins-spie2018}). Scanning FPIs on HIRMES ($25-122\,\mu\textrm{m}$) will be used to observe atomic and molecular lines in proto-planetary disks to study planet formation. In addition, HIRMES will map nearby galaxies in fine-structure lines to investigate the link between large scale galaxy properties and star forming regions as well as the properties of galactic centers \cite{hirmes,greg-hirmes-spie2018}. CCAT-prime will use scanning FPIs for one of its main science goals which is to study the epoch of reionization of the the universe via [CII] intensity mapping in the $750-1500\,\mu\textrm{m}$ regime \cite{ccat-prime,eve-ccatp-spie2018}

\section{DESIGN}
\label{sec:design}

\subsection{Metal Meshes}
\label{sec:design-metalmeshes}

Metal meshes have been used and studied as reflective filters and polarizers for microwave and infrared instruments since the mid-1900s \cite{ulrich_far-infrared_1967,renk_interference_1962}. Empirical studies have related the geometric properties of meshes to lumped-element circuit parameters to develop transmission line models of the meshes' optical transmittance, reflectance, and absorptance \cite{ulrich_effective_1967,ulrich_far-infrared_1967,marcuvitz_waveguide_1986}. Modern metal mesh design is primarily performed using software packages such as Computer Simulation Technology Microwave Studio (CST-MWS) or ANSYS High Frequency Structure Simulator (HFSS) which numerically solve Maxwell's equations for a given unit-cell of the structure \cite{ade_review_2006}.

\begin{figure} [ht]
\begin{center}
\begin{tabular}{c} 
\includegraphics[width=5in]{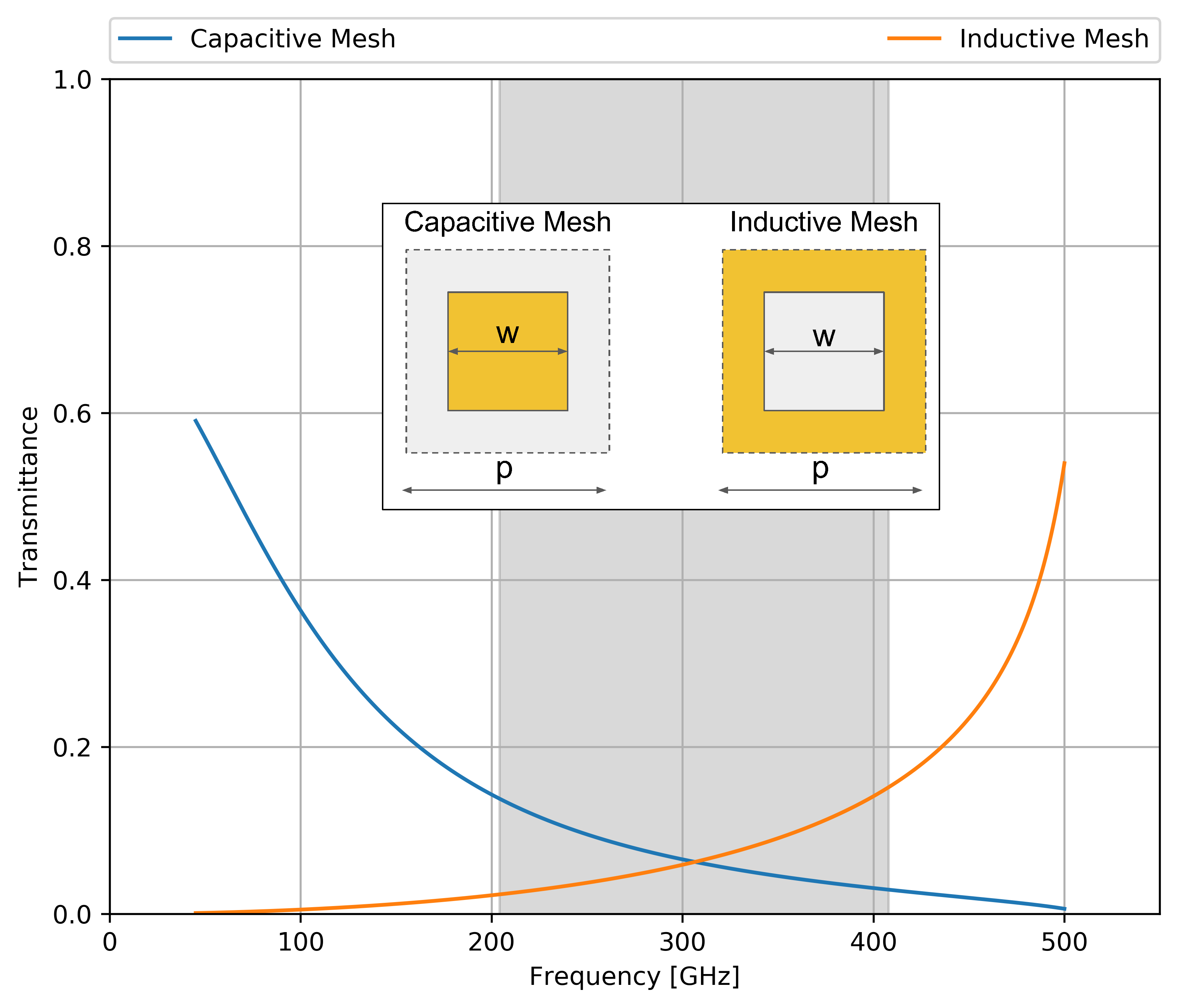}
\end{tabular}
\end{center}
\caption[example]{\label{fig:cst-mesh} 
Unit cell construction and CST-MWS transmittance simulation of inductive and capacitive meshes. The vertical grey band denotes the bandwidth of interest for a CCAT-prime etalon.}
\end{figure}

For this paper, we discuss the use of CST-MWS to study the transmission of a unit cell of inductive and capacitive gold meshes on silicon substrates at cryogenic temperatures. Figure \ref{fig:cst-mesh} shows the unit cell construction and the simulated transmittance of inductive and capacitive meshes on silicon. These simulations include the $1~K$ resistivity of gold ($\rho=2.2\times10^{-10},\Omega \cdot\mathrm{m}$) \cite{lide_crc_2016} and a 10 nm thick chromium adhesion layer ($\rho=1.25\times10^{7},\Omega \cdot\mathrm{m}$) between the silicon and the gold that is necessary for fabrication. The thickness of the gold layer is $500,\textrm{nm}$, much greater than the skin depth of gold in our frequency bands of interest for HIRMES and CCAT-prime. The silicon substrate is assumed to be a perfect dielectric ($\epsilon_r=11.7$) with no absorptive losses. The filter responses of both meshes are smooth for frequencies below the diffraction limit of the mesh. Diffraction occurs for wavelengths (in the index of silicon) smaller than the mesh's pitch. 

Most metal mesh filters can be modeled with capacitive and inductive mesh structures. By combining these two geometries, band-pass, band-stop, and multi-resonant filters can be constructed. For our purpose of wide bandwidth reflectors for FPIs, we will not combine both element types and we will restrict ourselves only to pure inductive or pure capacitive meshes whose reflectances are the most stable across wide bandwidths. In general, the filling ratio of a unit cell of the mesh determines the mesh's frequency-dependent transmittance and reflectance.

In order to build an efficient FPI for a wide bandwidth, it is important for the mirrors to have a roughly uniform reflectance of about $95\%$ across the bandwidth so that the etalon quality factor remains flat across the band. This is difficult to achieve over wide bandwidths using metal meshes because the capacitive and inductive filters tend towards zero or unity reflectance \cite{marcuvitz_waveguide_1986}. As an example for CCAT-prime, which requires an octave of bandwidth around 300 GHz, we are modeling FPIs which consist of one mirror with a capacitive mesh and another mirror with an inductive mesh. By combining the two mesh types, a flatter average reflectance is achieved over the bandwidth. Simulation results of FPIs using the capacitive-inductive design will be discussed below in Section \ref{sec:design-fpis}.

\subsection{Anti-Reflection Coatings}
\label{sec:design-arcs}

Silicon substrate mirrors for FPIs require very high-transmittance ARCs in order to mitigate Fresnel reflection by the un-metalized substrate surface. Any reflections in the substrate or ARC could cause parasitic resonances to leak into the FPI's transmission. Current techniques for millimeter wavelength applications involve cutting multi-layer pillar structures using dicing saw blades \cite{datta_large-aperture_2013}. Due to limitations on dicing saw blade thicknesses, this method is challenging to scale to sub-mm and far-IR wavelengths. In Section \ref{sec:fab-arcs} below, we discuss fabrication methods to etch metamaterial ARCs for sub-mm and far-IR wavelengths.

\begin{figure} [ht]
\begin{center}
\begin{tabular}{c} 
\includegraphics[width=5in]{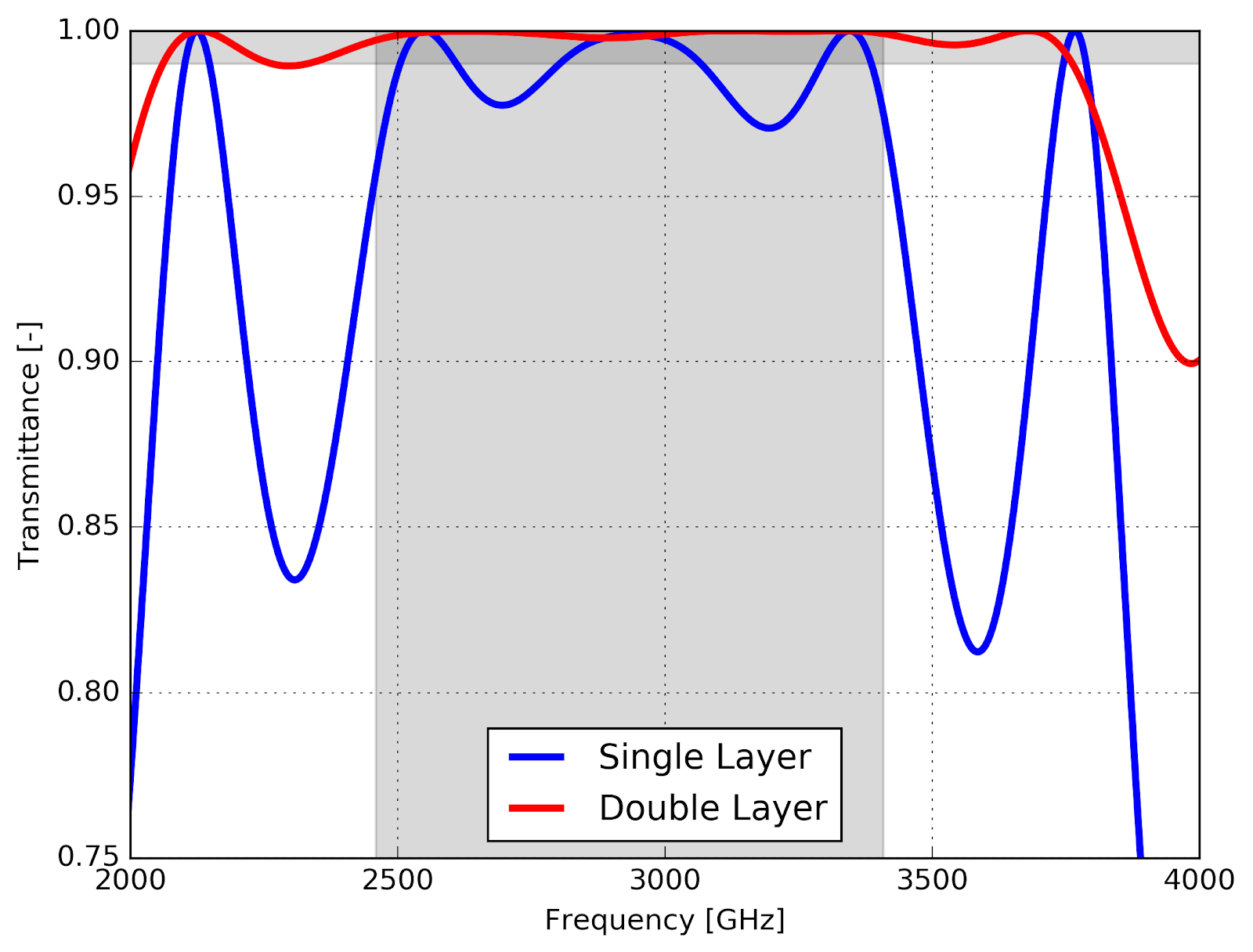}
\end{tabular}
\end{center}
\caption[example]{\label{fig:1larc-v-2larc} 
CST-MWS transmittance simulation of a single and double layer ARC. Both ARCs were designed for a center frequency of $2940\,\textrm{GHz}$. The vertical grey band denotes the bandwidth of interest for long wavelength HIRMES etalon. The horizontal grey band denotes 99\% transmittance.}
\end{figure}

The Fresnel reflection coefficient for normal incidence due to a thin film on a substrate can be shown to be $R=\frac{(n_1n_3-n_2^2)^2}{(n_1n_3+n_2^2)^2}$ where $n_1$, $n_2$, $n_3$ are the indices of refraction of the medium, film, and substrate respectively \cite{born_principles_1980}. Thus, the reflection of a silicon ($n_s=3.4$) substrate in vacuum ($n_0=1$) can be minimized with a film of index $n=\sqrt{3.4}$. Reflections are minimized by tuning the thickness of the film such that it is a quarter-wavelength (in the film index) of the frequencies of interest. This ensures that any reflections will destructively interfere with the incoming wave, effectively impedance matching the substrate and medium. This single layer will provide a narrow bandpass of nearly perfect transmission for wavelengths near the quarter-wavelength. The bandwidth of maximum transmittance can be widened with additional layers of stepped indices. Figure \ref{fig:1larc-v-2larc} shows the transmittance of single and double layer ARCs. The high-transmittance bandwidth is significantly widened with the additional layer.

Our multi-layered metamaterial ARC design consists of a grid of concentric, stacked, square holes or pillars.
The four-fold symmetry of a square grid has zero cross polarization at normal incidence \cite{mackay_proof_1989}. We estimate the index and thickness of each layer using a matrix formulation from the theory of layered dielectrics \cite{yeh_optical_1988}. The metamaterial geometry of each layer is then estimated using an effective capacitive circuit model \cite{biber_design_2003} for each metamaterial layer. The ARC is then simulated and optimized for the desired bandwidth in CST-MWS. In general, the index of refraction of a single metamaterial layer is a function of the area filling fraction of silicon and scales either linearly or quadratically depending on whether the geometry is holes or pillars \cite{jung-kubiak_antireflective_2017,defrance_1.6:1_2018}.


\subsection{Fabry-Perot Interferometers}
\label{sec:design-fpis}

\begin{figure} [ht]
\begin{center}
\begin{tabular}{c} 
\includegraphics[width=4in]{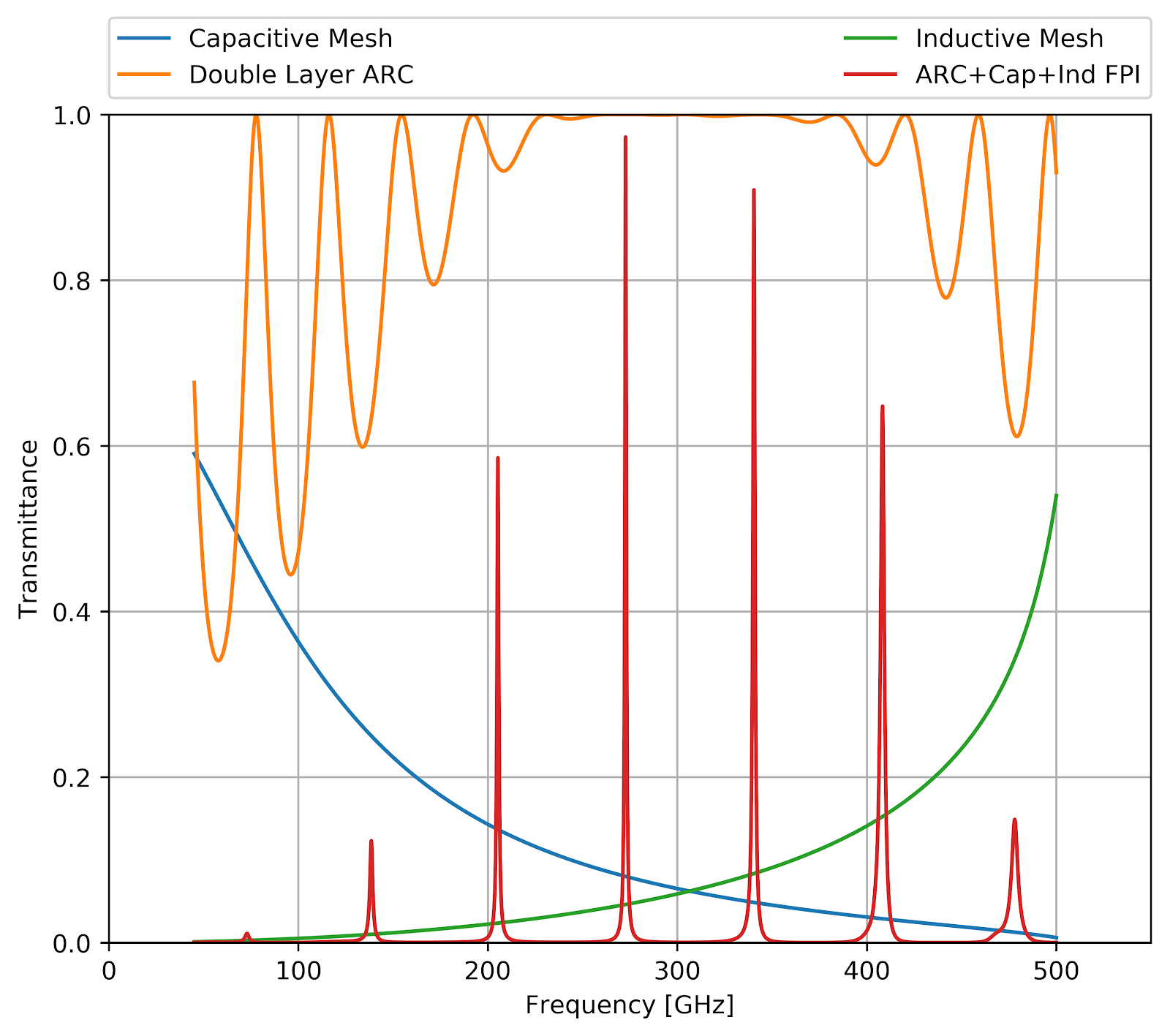}
\end{tabular}
\end{center}
\caption[example]{\label{fig:cst-fpi} 
CST-MWS transmittance simulation of inductive and capacitive meshes, two-layer ARC, and ARC-Si-mesh FPI model.}
\end{figure}

The full scanning FPI is simulated by combining our metal mesh and two-layer ARC models. The model consists of two $500\,\mu\textrm{m}$ silicon wafers held 2.2 mm apart, both with two-layer ARCs on their outer surfaces. On their inner surfaces, one wafer has a capacitive mesh and the other has an inductive mesh. Figure \ref{fig:cst-fpi} shows the transmittances of the individual meshes, the two-layer ARC, and the ARC-Si-mesh FPI. The ARC and metal mesh geometries used in Figure \ref{fig:cst-fpi} are given in Table \ref{tab:cst-params}. The FPI resonates for frequencies that are integer multiples of the cavity's fundamental frequency $f_0=\frac{c}{2nd\cos\theta}$ where $c$ is the speed of light, $n$ is the refractive index of the cavity, $d$ is the cavity spacing, and $\theta$ is the angle of incidence. The envelope of the resonances is caused by the frequency-dependent ARC and metal mesh filters being too reflective outside the band of interest. We explore the parameter space of metal mesh parameters in order to maximize the transmittance and flatten the finesse across wider bandwidths. 

We are also exploring the use of superconducting meshes instead of the traditional gold meshes. The finesse of an FPI can be thought of as the average number of bounces between the etalon mirrors that a resonance makes. At each reflection, ohmic losses decrease the resonance transmission. Free-standing meshes typically use low loss-metals, such as gold, to improve their transmission. By using a superconductor, ohmic losses can be further suppressed, and resonance transmission can be increased. The critical temperature of the superconductor will limit the performance of the mirror for incident wavelengths with energies that exceed the binding energy of the Cooper Pairs \cite{baryshev_progress_2011}


\begin{table}
\centering
\caption{ARC and metal mesh geometries used for CST-MWS simulations in Figure \ref{fig:cst-fpi}. A $500\,\mu\textrm{m}$ thick silicon substrate (perfect dielectric $\epsilon_r=11.7$) separated the ARC and metal mesh. A 10 nm thick chromium mesh layer with the same lateral dimensions of the gold meshes was placed between the substrate and the meshes in order to simulate the adhesion layer.} 
\label{tab:cst-params}
\begin{tabular}{|c|c|c|c|}
\hline
Layer                & Pitch ``p'' ($\mu\textrm{m}$) & Square size ``w'' ($\mu\textrm{m}$) & Thickness ($\mu\textrm{m}$) \\ \hline
ARC upper layer      & 137.3           & 127.2                 & 178.0          \\ \hline
ARC lower layer      & 137.3           & 85.5                  & 98.4           \\ \hline
Gold capacitive mesh & 171.3           & 164.4                 & 0.5            \\ \hline
Gold inductive mesh  & 171.3           & 101.1                 & 0.5            \\ \hline
\end{tabular}
\end{table}

\

\section{FABRICATION METHODS}
\label{sec:fabrication}

\subsection{Metal Meshes}
\label{sec:fab-metalmeshes}

We fabricate metal mesh filters using standard evaporation and lift-off lithography techniques. Negative lift-off resist (AZ nLOF 2020) is exposed with the mesh pattern using an ABM contact aligner and then developed. The patterned resist is descummed using an Anatech oxygen plasma barrel asher. Next, 10 nm of a chromium adhesion layer and 500 nm of gold are evaporated onto the patterned wafer using a CHA evaporator. The wafer is then placed in a bath of Microposit 1165 remover and left for 24 hours or until the photoresist is fully released. After rinsing and drying, the patterned metal mesh is microscopically examined for uniformity and correct dimensions. Using contact lithography, feature sizes of $\sim\!\!1\,\mu\textrm{m}$ can be achieved. Using stepper lithography, feature sizes of $\sim\!0.2\,\mu\textrm{m}$ can be achieved.

\subsection{Anti-Reflection Coatings}
\label{sec:fab-arcs}

\begin{figure} [ht]
\begin{center}
\begin{tabular}{c} 
\includegraphics[width=0.98\textwidth]{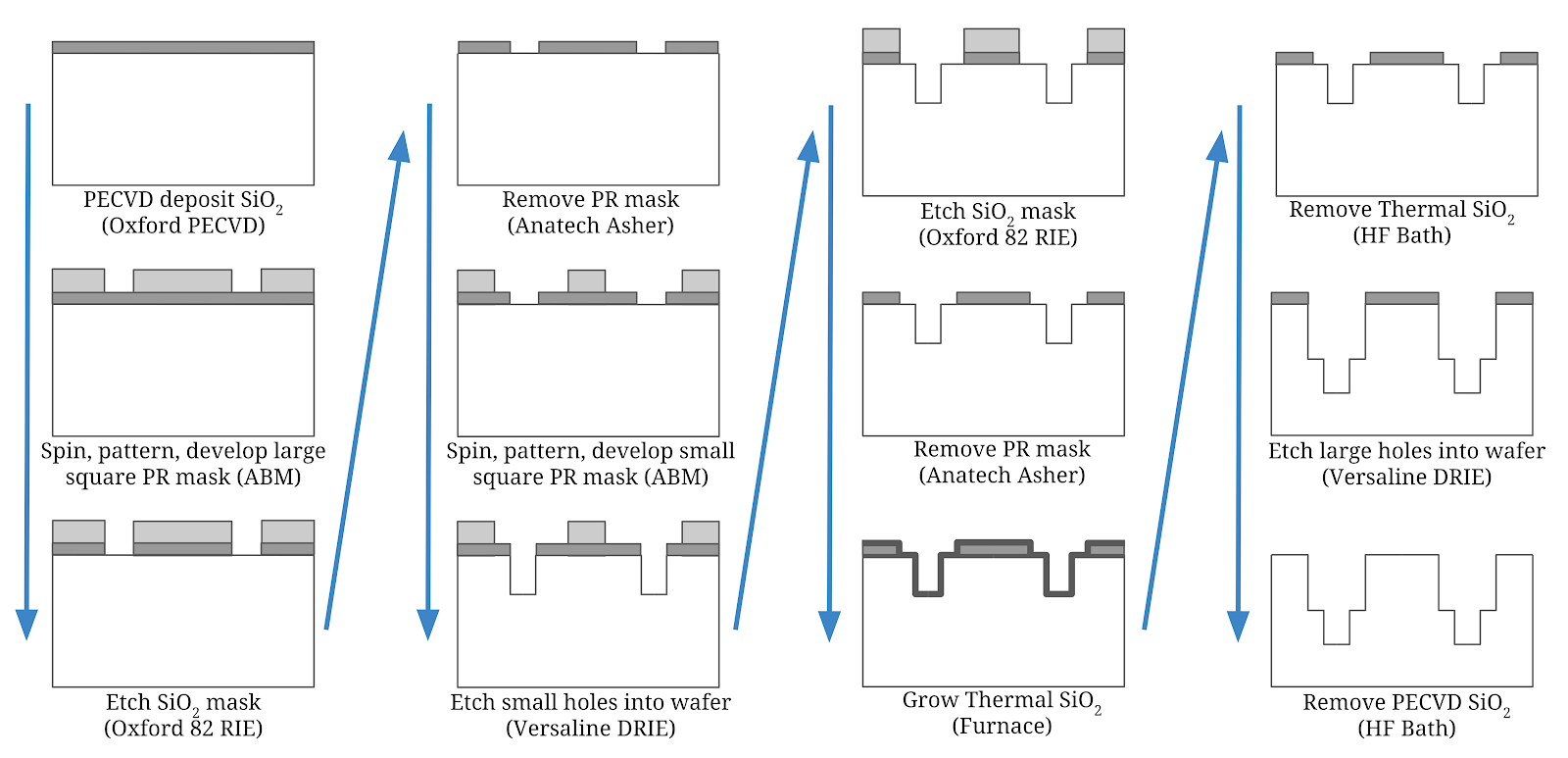}
\end{tabular}
\end{center}
\caption[example]{\label{fig:fab-arcs} 
Double layer metamaterial ARC recipe based on DRIE.}
\end{figure}

We fabricate metamaterial ARCs by using deep reactive ion etching (DRIE) to construct the desired geometry. DRIE is an anisotropic plasma etching technique used to etch silicon in the microfabrication industry. The process allows etching of high aspect ratio vertical structures to depths ranging from hundreds of nanometers to hundreds of microns. We etch metamaterial features into silicon wafers by applying etch masks to protect patterns of the silicon. For a single layer ARC, we pattern a square grid of photoresist which protects the silicon beneath it from being etched. For a double layer ARC, we stack silicon dioxide and photoresist etch masks of different lateral dimensions before etching any silicon. Figure \ref{fig:fab-arcs} outlines our recipe including the etch mask depositions and silicon etches.

The double layer recipe begins by depositing and patterning the silicon dioxide etch mask which defines the deepest metamaterial layer. A photoresist etch mask, which will define the uppermost metamaterial layer, is then patterned on top of the silicon dioxide. The lithography steps are performed using an ABM contact aligner or an ASML stepper, depending on the metamaterial feature sizes. The oxide is deposited using an Oxford plasma enhanced chemical vapor deposition (PECVD) tool. The lower metamaterial geometry is then etched using a PlasmaTherm Versaline DRIE. The oxide mask is then etched into a smaller etch mask using an Oxford reactive ion etch (RIE) tool and the photoresist etch mask is removed with an Anatech oxygen plasma barrel asher. 

After etching for a given amount of time, the wafer is removed from the DRIE tool and the etch depths are measured using a Zygo optical profilometer. After the profilometer measurements, the wafer is returned to the deep silicon etcher and etching continues. By performing these measurements throughout the full etch depth, the etch rate data can be acquired and used for precise control of the metamaterial layer thicknesses. In general, we have found that both metamaterial layers etch at different rates and as the layers become deeper, their etch rates slow. Characterizing the etch rates is key to successfully fabricating both metamaterial layers with their proper thicknesses. The Bosch process \cite{wu_high_2010} used in our DRIE tools can etch to $1/3\,\mu\textrm{m}$ depth resolution. The Zygo optical profilometer can measure depths with a resolution of up to 0.1 nm.

\begin{figure} [ht]
\begin{center}
\begin{tabular}{c} 
\includegraphics[width=0.98\textwidth]{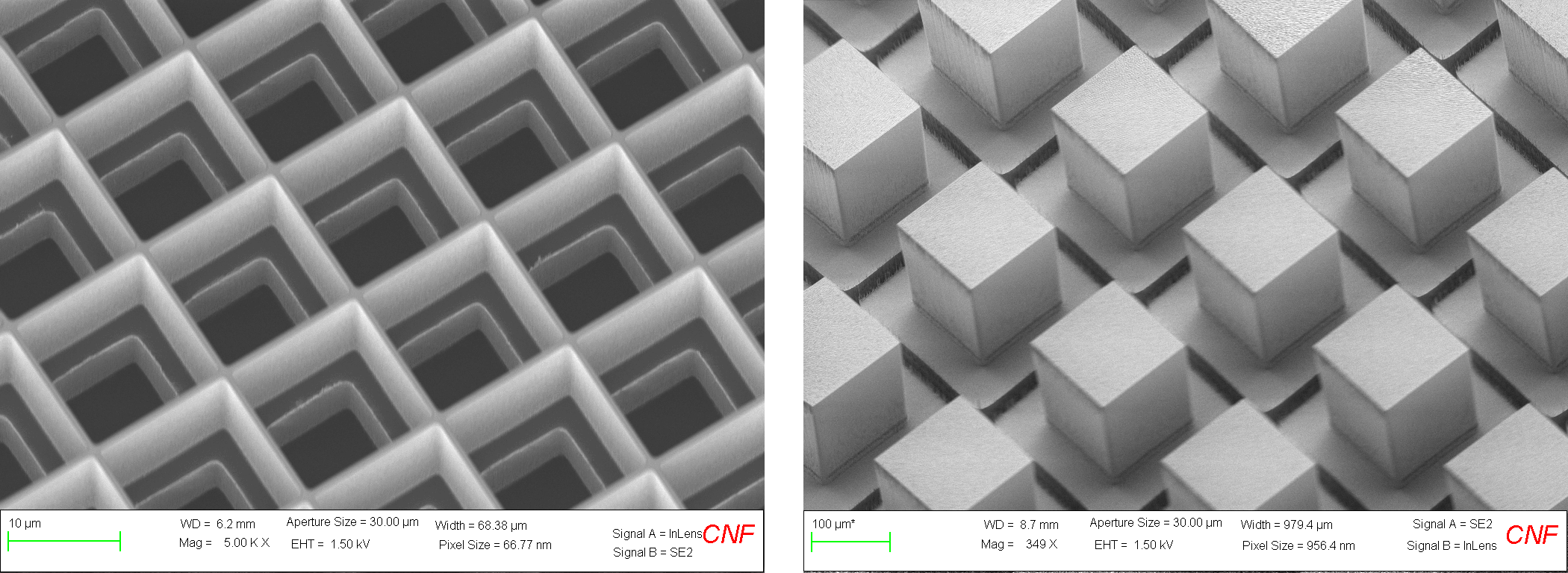}
\end{tabular}
\end{center}
\caption[example]{\label{fig:fab-arcs-sem} 
SEM micrographs. Left: Double metamterial layer coating using square hole geometry, designed for $102\,\mu\textrm{m}$ wavelengths. Right: Double metamaterial layer coating using square pillar geometry, designed for $1\,\textrm{mm}$ wavelengths.}
\end{figure}

To clean the surface geometry and remove any sidewall passivation left from the first DRIE step, the wafer surface is oxidized and etched using a buffered oxide etch of hydrofluoric acid. The upper metamaterial layer is then etched into the silicon and the remaining silicon dioxide etch mask is removed. Figure \ref{fig:fab-arcs-sem} shows two micrographs taken with a scanning electron microscope of double layer metamaterial coatings. The left image is a geometry of concentric square holes with a pitch of $14\,\mu\textrm{m}$, designed for a band centered at $102\,\mu\textrm{m}$. The right image is a geometry of concentric square pillars with a pitch of $171\,\mu\textrm{m}$, designed for a band centered at $1\,\textrm{mm}$.

\section{FUTURE WORK}
\label{sec:future}
We are currently fabricating double layer ARCs and metal meshes on high-resistivity ($\rho>10,000 \,\Omega\cdot \textrm{cm}$) silicon wafers. Their optical performance will be measured using a Fourier transform spectrometer (FTS) and compared to CST-MWS simulations. The results will be used to iterate on the fabrication procedures and models until measurements and simulations are in agreement. Single FPI mirrors will be fabricated by etching one side of a high-resistivity wafer with a double layer ARC and patterning the other side with a metal mesh. 

As an intermediate step in our fabrication process and to verify our modeling, we will fabricate a fixed FPI device by patterning metal meshes on either side of a single high-resistivity silicon wafer, using the wafer as the resonance cavity. FTS measurements of this sample will help to characterize the metal mesh and silicon substrate performance. This device can also be used as a wavelength calibrator for other instruments since it effectively acts as a multi-band narrow pass filter. 

Modeling and fabrication techniques for superconducting meshes are in development. Samples will be fabricated and characterized using a cryogenic FTS. Antireflection coated mirrors and fixed FPIs will be fabricated and tested. We anticipate measuring decreased ohmic loss and increased FPI transmittance.

\section{CONCLUSION}
\label{sec:conclusion}
Fabrication of metal mesh reflectors and anti-reflection coated silicon surfaces have been demonstrated. In the near future we combine the two to create high efficiency, broad-band mirrors for astrophysics. These fabrication techniques are shown to be feasible for a large range of bandwidths from the far-IR to the millimeter. Fabrication and measurements of metal mesh and anti-reflection coated silicon substrate mirrors is upcoming.

The silicon-substrate based mirrors that are developed here will be used in upcoming scanning FPI instruments such as Prime-Cam in CCAT-prime and HIRMES in SOFIA. Other optical elements that could utilize these modeling and fabrication techniques include silicon substrate metal mesh band-defining filters, anti-reflection coated silicon lenses, metamaterial lenses, half-wave plates, or metamaterial dielectric high-reflectance mirrors.

\acknowledgements
Work by NFC and BJK was supported by a NASA Space Technology Research Fellowship.
Fabrication work at the Cornell NanoScale Science and Technology Facility supported by NASA Grant NNX16AC72G.
MDN acknowledges support from NSF award AST-1454881.
This work was performed in part at Cornell NanoScale Facility, an NNCI member supported by NSF Grant ECCS-1542081.


\bibliography{SPIE2018,report} 

\begin{thebibliography}{10}

\bibitem{lacy-thesis}
{Lacy}, J.~H., {\em {Infrared fine-structure line emission from the galactic
  center}}, PhD thesis, UC Berkeley (1982).

\bibitem{bradford}
{Bradford, C.~M., Stacey, G.~J., Swain, M.~R., et al.}, ``{SPIFI: a
  direct-detection imaging spectrometer for submillimeter wavelengths},'' {\em
  Appl. Optics}~{\bf 41},  2561--2574 (2002).

\bibitem{watson-thesis}
{Watson}, D.~M., {\em {Shockwave and Mass Outflow in the Orion Molecular Cloud:
  Observations of Far-Infrared Emission Lines of Carbon Monoxide, Hydroxyl, and
  Ammonia}}, PhD thesis, UC Berkeley (1982).

\bibitem{poglitsch}
{Poglitsch, A., Beeman, J.~W., Geis, N., et al.}, ``{The MPE/UCB far-infrared
  imaging Fabry-Perot interferometer (FIFI)},'' {\em International Journal of
  Infrared and Millimeter Waves}~{\bf 12},  859--884 (1991).

\bibitem{latvakoski}
{Latvakoski, H.~M., Stacey, G.~J., Gull, G.~E., et al.}, ``Kuiper widefield
  infrared camera far-infrared imaging of the galactic center: The
  circumnuclear disk revealed,'' {\em Ap. J.}~{\bf 511},  761--773 (1999).

\bibitem{clegg}
{Clegg, P.~E., et. al}, ``{The ISO Long-Wavelength Spectrometer},'' {\em
  Astron. Astrophys.}~{\bf 315},  L38--L42 (2002).

\bibitem{datta_large-aperture_2013}
Datta, R., Munson, C.~D., Niemack, M.~D., McMahon, J.~J., Britton, J., Wollack,
  E.~J., Beall, J., Devlin, M.~J., Fowler, J., Gallardo, P., Hubmayr, J.,
  Irwin, K., Newburgh, L., Nibarger, J.~P., Page, L., Quijada, M.~A., Schmitt,
  B.~L., Staggs, S.~T., Thornton, R., and Zhang, L., ``Large-aperture
  wide-bandwidth antireflection-coated silicon lenses for millimeter
  wavelengths,'' {\em Applied Optics}~{\bf 52},  8747--8758 (Dec. 2013).

\bibitem{gallardo_deep_2016}
Gallardo, P.~A., Koopman, B.~J., Cothard, N., Bruno, S. M.~M., Cortes-Medellin,
  G., Marchetti, G., Miller, K.~H., Mockler, B., Niemack, M.~D., Stacey, G.,
  and {others}, ``Deep {Reactive} {Ion} {Etched} {Anti}-{Reflection} {Coatings}
  for {Sub}-millimeter {Silicon} {Optics},'' {\em arXiv preprint
  arXiv:1610.07655}  (2016).

\bibitem{ccat-prime}
{Stacey}, G.~J., ``{CCAT-Prime: an ultra-widefield submillimeter observatory on
  Cerro Chajnantor},'' in [{\em Millimeter, Submillimeter, and Far-Infrared
  Detectors and Instrumentation for Astronomy IX}{\nolinebreak\hspace{0.1em}]},
   {\em Proc. SPIE},  10700--53 (in Press 2018).

\bibitem{eve-ccatp-spie2018}
{Vavagiakis}, E.~M., ``{Prime-Cam: A first-light instrument for the CCAT-prime
  Telescope},'' in [{\em Millimeter, Submillimeter, and Far-Infrared Detectors
  and Instrumentation for Astronomy IX}{\nolinebreak\hspace{0.1em}]},  {\em
  Proc. SPIE},  10708--64 (in Press 2018).

\bibitem{hirmes}
{Kutyrev}, A.~S., ``{HIRMES: the third generation instrument for SOFIA},'' in
  [{\em Millimeter, Submillimeter, and Far-Infrared Detectors and
  Instrumentation for Astronomy IX}{\nolinebreak\hspace{0.1em}]},  {\em Proc.
  SPIE},  10708--22 (in Press 2018).

\bibitem{greg-hirmes-spie2018}
{Douthit}, J.~G., ``{Development of the Fabry-Perot interferometers for the
  HIRMES spectrometer on SOFIA},'' in [{\em Millimeter, Submillimeter, and
  Far-Infrared Detectors and Instrumentation for Astronomy
  IX}{\nolinebreak\hspace{0.1em}]},  {\em Proc. SPIE},  10708--59 (in Press
  2018).

\bibitem{origins-spie2018}
{Leisawitz}, D.~T., {Amatucci}, E.~G., and {Carter}, R.~C., ``{The Origins
  Space Telescope: mission concept overview},'' in [{\em Space Telescopes and
  Instrumentation 2018: Optical, Infrared, and Millimeter
  Wave}{\nolinebreak\hspace{0.1em}]},  {\em Proc. SPIE},  10698--40 (in Press
  2018).

\bibitem{ulrich_far-infrared_1967}
Ulrich, R., ``Far-infrared properties of metallic mesh and its complementary
  structure,'' {\em Infrared Physics}~{\bf 7}(1),  37--55 (1967).

\bibitem{renk_interference_1962}
Renk, K.~F. and Genzel, L., ``Interference filters and {Fabry}-{Perot}
  interferometers for the far infrared,'' {\em Applied Optics}~{\bf 1}(5),
  643--648 (1962).

\bibitem{ulrich_effective_1967}
Ulrich, R., ``Effective low-pass filters for far infrared frequencies,'' {\em
  Infrared Physics}~{\bf 7},  65--74 (June 1967).

\bibitem{marcuvitz_waveguide_1986}
Marcuvitz, N.,  [{\em Waveguide handbook}{\nolinebreak\hspace{0.1em}]}, P.
  Peregrinus on behalf of the Institution of Electrical Engineers, London, UK
  (1986).

\bibitem{ade_review_2006}
Ade, P. A.~R., Pisano, G., Tucker, C., and Weaver, S., ``A review of metal mesh
  filters,''  {\bf 6275},  62750U--62750U--15 (2006).

\bibitem{lide_crc_2016}
Haynes, W.~M., Bruno, T.~J., and Lide, D.~R., eds.,  [{\em {CRC} handbook of
  chemistry and physics: a ready-reference book of chemical and physical
  data}{\nolinebreak\hspace{0.1em}]}, CRC Press, Boca Raton, 97th edition.~ed.
  (2016).

\bibitem{born_principles_1980}
Born, M.,  [{\em Principles of optics: electromagnetic theory of propagation,
  interference and diffraction of light}{\nolinebreak\hspace{0.1em}]}, Pergamon
  Press, Oxford, 6th ed.~ed. (1980).

\bibitem{mackay_proof_1989}
Mackay, A., ``Proof of polarisation independence and nonexistence of crosspolar
  terms for targets presenting n-fold (n{\textgreater}2) rotational symmetry
  with special reference to frequency-selective surfaces,'' {\em Electronics
  Letters}~{\bf 25},  1624--1625 (Nov. 1989).

\bibitem{yeh_optical_1988}
Yeh, P.,  [{\em Optical waves in layered media}{\nolinebreak\hspace{0.1em}]},
  Wiley, New York (1988).

\bibitem{biber_design_2003}
Biber, S., Richter, J., Martius, S., and Schmidt, L.~P., ``Design of
  {Artificial} {Dielectrics} for {Anti}-{Reflection}-{Coatings},'' in [{\em
  2003 33rd {European} {Microwave} {Conference}}{\nolinebreak\hspace{0.1em}]},
   1115--1118 (Oct. 2003).

\bibitem{jung-kubiak_antireflective_2017}
Jung-Kubiak, C., Sayers, J., Hollister, M.~I., Bose, A., Yoshida, H., Liao, L.,
  Wong, J., Radford, S., Chattopadhyay, G., and Golwala, S., ``Antireflective
  textured silicon optics at millimeter and submillimeter wavelengths,'' in
  [{\em 2017 11th {European} {Conference} on {Antennas} and {Propagation}
  ({EUCAP})}{\nolinebreak\hspace{0.1em}]},   959--961 (Mar. 2017).

\bibitem{defrance_1.6:1_2018}
Defrance, F., Jung-Kubiak, C., Sayers, J., Connors, J., deYoung, C., Hollister,
  M.~I., Yoshida, H., Chattopadhyay, G., Golwala, S.~R., and Radford, S. J.~E.,
  ``1.6:1 bandwidth two-layer antireflection structure for silicon matched to
  the 190-310 {GHz} atmospheric window,'' {\em Applied Optics}~{\bf 57},
  5196--5209 (June 2018).

\bibitem{baryshev_progress_2011}
Baryshev, A., Baselmans, J. J.~A., Freni, A., Gerini, G., Hoevers, H., Iacono,
  A., and Neto, A., ``Progress in {Antenna} {Coupled} {Kinetic} {Inductance}
  {Detectors},'' {\em IEEE Transactions on Terahertz Science and
  Technology}~{\bf 1},  112--123 (Sept. 2011).

\bibitem{wu_high_2010}
Wu, B., Kumar, A., and Pamarthy, S., ``High aspect ratio silicon etch: {A}
  review,'' {\em Journal of Applied Physics}~{\bf 108},  051101 (Sept. 2010).

\end{thebibliography}
\bibliographystyle{spiebib} 

\end{document}